# Verifying Safety-Critical Timing and Memory-Usage Properties of Embedded Software by Abstract Interpretation

Reinhold Heckmann, Christian Ferdinand
AbsInt Angewandte Informatik GmbH
Stuhlsatzenhausweg 69; 66123 Saarbrücken, Germany
info@absint.com

**Abstract**

*Static program analysis by abstract interpretation is an efficient method to determine properties of embedded software. One example is value analysis, which determines the values stored in the processor registers. Its results are used as input to more advanced analyses, which ultimately yield information about the stack usage and the timing behavior of embedded software.*

## 1. Abstract Interpretation

Failure of a safety-critical application on an embedded processor can lead to severe damage or even loss of life. Therefore, utmost carefulness and state-of-the-art machinery have to be applied to make sure that an application meets all requirements. To do so lies in the responsibility of the system designer(s).

Classical software validation methods like code review and testing with debugging are very expensive. Furthermore, they cannot really guarantee the absence of errors. In contrast, *abstract interpretation* [1] is a formal verification method that yields statements valid for all program runs with all inputs, e.g., absence of violations of timing or space constraints, or absence of runtime errors.

Nowadays tools based on abstract interpretation are commercially available and have proved their usability in industrial practice. For example, stack overflow can be detected by **AbsInt**'s **StackAnalyzer**, and violations of timing constraints are found by **AbsInt**'s **aiT** tool [2] that determines upper bounds for the worst-case execution times of the tasks of an application. Among other things, these tools perform a *value analysis* that tries to determine the values stored in the processor's memory for every program point.

Value analysis is a static analysis method based on abstract interpretation. It produces results valid for every program run and all inputs to the program. Therefore, it cannot always predict an exact value for a memory location, but determines *abstract values* instead that stand for sets of concrete values. There are several variants of value analysis depending on what kinds of abstract values are used. In *constant propagation*, an abstract value is either a single concrete value or the statement that no information about the value is known. In *interval analysis*, abstract values are intervals that are guaranteed to contain the exact values. Further extensions of value analysis record known equalities between otherwise unknown values, or more generally, upper and lower bounds for their differences, or even more generally, arbitrary linear constraints between values.

Value analysis, even in its simple form as interval analysis, has various applications as an auxiliary method providing input for other analysis tasks. Some of these applications are listed in the next few sections.

## 2. Stack usage analysis

A possible cause of catastrophic failure is stack overflow that usually leads to run-time errors that are difficult to diagnose. The problem is that the memory area for the stack usually must be reserved by the programmer. Underestimation of the maximum stack usage leads to stack overflow, while overestimation means wasting memory resources. Measuring the maximum stack usage with a debugger is no solution since one only obtains results for single program runs with fixed inputs. Even repeated measurements cannot guarantee that the maximum stack usage is ever observed.

**AbsInt**'s **StackAnalyzer** provides a solution to this problem: By concentrating on the value of the stack pointer during value analysis, the tool can figure out how the stack increases and decreases along the various control-flow paths. The predicted worst-case stack usages of individual tasks in a system can be used in an automated overall stack usage analysis for all tasks running on one Electronic Control Unit, as described in [3] for systems managed by an OSEK/VDX real-time operating system.

## 3. Worst-case execution time prediction

Many tasks in safety-critical embedded systems have hard real-time characteristics. Failure to meet deadlines may be as harmful as producing wrong output or failure to work at all. Yet the determination of the Worst-Case Execution Time (WCET) of a task is a difficult problem because of the characteristics of modern software and hardware.

Embedded control software (e.g., in the automotive industries) tends to be large and complex. The software in a single electronic control unit is usually developed by several people, several groups or even several different providers. It is typically combined with third-party software such as real-time operating systems and/or communication libraries.

Caches and branch target buffers are used in virtually all performance-oriented processors to reduce the number of accesses to slow memory. Pipelines enable acceleration by overlapping the executions of different instructions. Consequently the timing of the instructions depends on the execution history.

The widely used classical methods of predicting execution times are not generally applicable. Software monitoring and dual-loop benchmark change the code, which in turn changes the cache behavior. Hardware simulation, emulation, or direct measurement with logic analyzers can only determine the execution time for some fixed inputs.

In contrast, abstract interpretation can be used to efficiently compute a safe approximation for all possible cache and pipeline states that can occur at a program point in any program run with any input. These results can be combined with ILP (Integer Linear Programming) techniques to safely predict the worst-case execution time and a corresponding worst-case execution path.

**AbsInt**'s WCET tool **aiT** determines the WCET of a program task in several phases [2]: **CFG building** decodes, i.e. identifies instructions, and reconstructs the control-flow graph (CFG) from a binary program; **value analysis** computes value ranges for registers and address ranges for instructions accessing memory; **loop bound analysis** determines upper bounds for the number of iterations of simple loops; **cache analysis** classifies memory references as cache misses or hits; **pipeline analysis** predicts the behavior of the program on the processor pipeline; **path analysis** determines a worst-case execution path of the program.

The results of value analysis are used to determine possible addresses of indirect memory accesses—important for cache analysis—and in loop bound analysis. They are usually so good that only a few indirect accesses cannot be determined exactly. Value analysis can also determine that certain conditions always evaluate to true or always evaluate to false. As a consequence, certain paths controlled by such conditions are never executed. Therefore, their execution time does not contribute to the overall WCET of the program, and need not be determined in the first place.

Cache Analysis uses the results of value analysis to predict the behavior of the (data) cache. The results of cache analysis are used within pipeline analysis allowing the prediction of pipeline stalls due to cache misses. The combined results of the cache and pipeline analyses are the basis for computing the execution times of program paths. Separating WCET determination into several phases makes it possible to use different methods tailored to the subtasks. Value analysis, cache analysis, and pipeline analysis are done by abstract interpretation [1]. Integer linear programming is used for path analysis.

**aiT** allows to inspect the timing behavior of (time-critical parts of) program tasks. The analysis results are determined without the need to change the code and hold for all executions. **aiT** takes into account the combination of all the different hardware characteristics while still obtaining tight upper bounds for the WCET of a given program in reasonable time. Its results are documented in a report file and as annotations in the control-flow graph that can be visualized using **AbsInt**'s graph viewer **aiSee**.

## 4. Conclusion

Tools based on abstract interpretation can perform static program analysis of embedded applications. Their results hold for all program runs with arbitrary inputs. Employing static analyzers is thus orthogonal to classical testing, which yields very precise results, but only for selected program runs with specific inputs. The usage of static analyzers enables one to develop complex systems on state-of-the-art hardware, increases safety, and saves development time. Precise stack usage and timing predictions enable the most cost-efficient hardware to be chosen. As recent trends, e.g., in automotive industries (X-by-wire, time-triggered protocols) require knowledge on the WCETs of tasks, a tool like **aiT** is of high importance.